\documentclass[aps,twocolumn,pre,floatfix]{revtex4-2}

\usepackage{xcolor,hyperref}
\hypersetup{
   colorlinks,
   linkcolor={blue!50!black},
   citecolor={blue!50!black},
   urlcolor={blue!80!black}
}

\usepackage{epsfig, amsmath}
\usepackage{bm}
\usepackage{soul,ulem}
\usepackage{hyperref}
\usepackage{footmisc}
\usepackage{wrapfig}
\DeclareMathAlphabet{\mathitbf}{OML}{cmm}{b}{it}
\newcommand{\tripleCdot}{:\!\cdot\,}

\newcommand{\xv}{\mathitbf x}

\newcommand{\piv}{\bm{\pi}}
\newcommand{\psiv}{\bm{\psi}}

\newcommand{\calBold}[1]{\mbox{\boldmath${\cal #1}$}}
\newcommand{\dbar}{{\,\mathchar'26\mkern-12mu d}}

\newcommand{\mathBold}[1]{\mbox{\boldmath$#1$}}

\begin{document}

\title{Ultra-high Poisson's ratio glasses}
\author{Edan Lerner}
\email{e.lerner@uva.nl}
\affiliation{Institute for Theoretical Physics, University of Amsterdam, Science Park 904, 1098 XH Amsterdam, the Netherlands}

\begin{abstract}
    The manner in which metallic glasses fail under external loading is known to correlate well with those glasses' Poisson's ratio $\nu$: low-$\nu$ (compressible) glasses typically feature brittle failure patterns with scarce plastic deformation, while high-$\nu$ (incompressible) glasses typically fail in a ductile manner, accompanied by a high degree of plastic deformation and extensive liquid-like flow. Since the technological utility of metallic glasses depends on their ductility, materials scientists have been concerned with fabricating high-$\nu$ glassy alloys. To shed light on the underlying micromechanical origin of high-$\nu$ metallic glasses, we employ computer simulations of a simple glass-forming model with a single tunable parameter that controls the interparticle-potential's stiffness. We show that the presented model gives rise to ultra high-$\nu$ glasses, reaching $\nu\!=\!0.45$ and thus exceeding the most incompressible laboratory metallic glass. We discuss the possible role of the so-called unjamming transition in controlling the elasticity of ultra high-$\nu$ glasses. To this aim, we show that our higher-$\nu$ computer glasses host relatively softer quasilocalized glassy excitations, and establish relations between their associated characteristic frequency, macroscopic elasticity, and mechanical disorder.
\end{abstract}

\maketitle

\section{Introduction}
\label{sec:introdution}

The Poisson's ratio (PR) $\nu$ of a material represents the extent to which an applied uniaxial load results in deformation in the plane perpendicular to the loading axis. In three dimensions, an isotropic material is deemed \emph{incompressible} if it features $\nu\!=\!1/2$, then a uniaxial load will result in an elastic (reversible) deformation that conserves the material's volume perfectly~\cite{landau_lifshitz_elasticity}. Some soft, disordered materials such as rubber are nearly incompressible, featuring $\nu\!\approx\!1/2$, while e.g.~cork is highly compressible, with a PR that is close to zero~\cite{cork_1947}. In anisotropic, perfect-crystalline materials, the PR can span a much larger range~\cite{extreme_nu_acta_2010}, as is also the case for auxetic materials~\cite{review_auxetic_1} and metamaterials~\cite{Ren_2018}.

The PR is an important material parameter; despite being defined based solely \emph{linear}-elastic moduli, it can serve as an indicator of a material's nonlinear, dissipative mechanical response~\cite{Greer_2005,Greaves2011}, structural relaxation patterns~\cite{Novikov2004,Yannopoulos2006}, and vibrational properties~\cite{Duval_2013}. For instance, several investigations of the fracture toughness of metallic glasses have pointed out that it is highly correlated with the PR of those materials: metallic glasses featuring higher PRs are observed to undergo a sharp brittle-to-ductile crossover in their nonlinear mechanical response~\cite{Greer_2005,Greaves2011,Poon_2008,shi_2014_intrinsic_ductility,wang_2014}. While the significance of the correlation between metallic glasses' PRs and their failure patterns is still debated~\cite{Rycroft2012,david_fracture_2021}, understanding the mechanism that controls a material's PR remains an interesting challenge~\cite{wang_2012,experimental_inannealability_AM_2016}. In particular, different from polymeric, rubber-like materials in which elasticity is entropic in nature, the elastic properties of metallic glasses should be understood predominantly in terms of the interaction potentials between those materials' constituent particles, and their associated emergent micromechanics. 

Since the robustness of metallic glasses against catastrophic failure is important for their technological utility, designing ductile metallic glasses with high PRs is a burning challenge~\cite{old_metallic_glass_high_pr,jan_2004_prl,Demetriou2011,ultra_high_PR_Demetriou_2011,strengthening_metallic_glasses_2014}. To date, the highest-PR metallic glass --- a platinum-rich alloy --- was fabricated by Demetriou et al.~and reported upon in Ref.~\cite{ultra_high_PR_Demetriou_2011}; it features a PR of 0.43, and is therefore referred to in~\cite{ultra_high_PR_Demetriou_2011} as `liquid-like'. In addition, the same material is one of the \emph{stiffest} metallic glasses recorded to date, boasting a bulk modulus of $K\!\approx\!217\, \mbox{GPa}$.

What microscopic mechanism is responsible for these ultra-high PR materials nearly incompressible character? One intriguing model of an elastic solid that can be driven to the incompressible-limit is soft-sphere packings~\cite{ohern2003}. These systems of purely \emph{repulsive}, frictionless spheres are typically studied at vanishing pressures, such that they reside close to the \emph{unjamming transition}~\cite{liu_review,van_hecke_review} -- the point at which the relative shear rigidity $G/K\!\to\!0$ (with $G$ denoting the shear modulus) under slow decompression, and consequently $\nu\!\to\!1/2$ (and recall that $\nu\!\equiv\!(3\!-\!2G/K)/(6\!+\!2G/K)$). In this work we discuss and test the hypothesis that a similar mechanism that gives rise to incompressibility in soft-sphere packings near the unjamming point -- is also present in the high-PR glassy alloys studied in Refs.~\cite{jan_2004_prl,Demetriou2011,ultra_high_PR_Demetriou_2011}. This hypothesis is supported by the high bulk moduli featured by those high-PR metallic glasses; these suggest that introducing \emph{stiffer} repulsive interaction potentials between glasses' constituent particles would enhance steric exclusion effects in the spatial organization of those glasses' microstructures, thus inducing unjamming-like mechanics, and high emergent PRs of those glasses.

To test this hypothesis, we employ a simple glass forming model whose particles interact via a pairwise potential in which the stiffness associated with the short-range repulsion can be tuned via a single parameter (denoted Q, cf.~Fig.~\ref{fig:potential} below). Using this model, we create glasses --- with finite tensile yield-stresses \emph{and} at vanishing pressures --- that feature ultra-high PRs, higher than the current highest-PR laboratory metallic glass~\cite{ultra_high_PR_Demetriou_2011}. We discuss indications supporting that an unjamming-like mechanism is responsible for the ultra-high PRs of our computer glasses. This is done by investigating the structural properties and mechanical disorder of our ultra-high PR glasses and showing that they feature an abundance of very soft glassy defects, whose relative characteristic frequency decreases with increasing PR. We consider in addition a measure of mesoscopic mechanical disorder (see precise definitions below), and establish a previously proposed relation between it and the density of glassy excitations. Finally, we highlight an intriguing difference between the micromechanics of our ultra-high PR glasses and of soft-sphere packings near unjamming. All in all, our results demonstrate that unjamming phenomenology could be highly relevant for understanding the elastic properties of some atomistic, laboratory structural glasses.


\section{Model and methods}
\label{sec:models_and_methods}
\begin{figure}[ht!]
  \includegraphics[width = 0.5\textwidth]{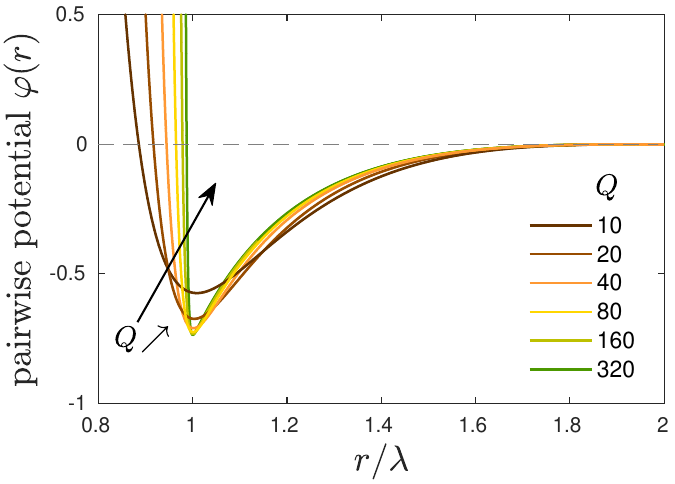}
  \caption{\footnotesize Pairwise potential $\varphi(r)$, expressed in terms of the microscopic units of energy $\varepsilon$, plotted for the studied values of the parameter $Q$ that controls the interparticle stiffness.}
  \label{fig:potential}
\end{figure}

We employ a 50:50 binary mixture of `large' and `small' particles of equal mass $m$ interacting via the pairwise potential
\begin{equation}
    \label{eq:pairwise_potential}
    \varphi(r_{ij}) = \frac{4\varepsilon}{Q-4}\left[ \left(\frac{\lambda_{ij}}{r_{ij}}\right)^{\!\!Q} - \frac{Q}{4} \left(\frac{\lambda_{ij}}{r_{ij}}\right)^4 + \tilde{\varphi}_{\mbox{\tiny smooth}}(r_{ij})\right]\,, 
\end{equation}
for $\frac{r_{ij}}{\lambda_{ij}}<x_{\rm c}$, and $\varphi(r_{ij})\!=\!0$ otherwise. Here $\varepsilon$ represents a microscopic energy scale, $\lambda_{ij}\!=\!\lambda, 1.18\lambda,$ or $1.4\lambda$ for `small'-`small', `small-large' or `large-large' interactions, respectively, $\lambda$ forms the microscopic units of length, and $x_{\rm c}$ is the dimensionless cutoff distance. The function $\tilde{\varphi}_{\mbox{\tiny smooth}}(r_{ij})$ is defined as
\begin{equation}
    \tilde{\varphi}_{\mbox{\tiny smooth}}(r_{ij})\!=\!c_4\left(\frac{r_{ij}}{\lambda_{ij}}\right)^4\!+\!c_2\left(\frac{r_{ij}}{\lambda_{ij}}\right)^2\!+\!c_0\,.
\end{equation}
The coefficients $c_4,c_2,c_0$ are functions of $Q$ and of the dimensionless cutoff distance $x_{\rm c}$, determined as to ensure that the potential and two of its derivatives vanish continuously at $x_{\rm c}$, leading to
\begin{eqnarray}
c_4 & = & -Q\big( (Q+2)x_{\rm c}^{-(Q+4)} - 6x_{\rm c}^{-8} \big)/(2Q-8)\,, \nonumber \\
c_2 & = & Q\big( (Q+4)x_{\rm c}^{-(Q+2)} - 8x_{\rm c}^{-6} \big)/(Q-4)\,, \nonumber \\
c_0 & = & -\big((Q+2)(Q+4)x_{\rm c}^{-Q} - 12Qx_{\rm c}^{-4}\big)/(2Q-8)\,. \nonumber
\end{eqnarray}
The pairwise potential $\varphi(r)$ is presented in Fig.~\ref{fig:potential} for $x_{\rm c}\!=\!2$ and various values of $Q$ as detailed by the figure legend.


In order to mimic laboratory glasses at ambient pressures, we chose the dimensionless densities $\lambda^3 N/V\!=\!0.777, 0.737, 0.708, 0.694, 0.6802, 0.677$ and 0.675 for $Q\!=\!10,20,40,80,160,240$ and 320, respectively, where $N$ denotes the number of particles and $V$ the simulation-box volume. The resulting mean pressure to bulk-modulus ratio $p/K$ of our glasses is on the order of $10^{-3}$ or smaller for all of our glass ensembles, as shown in Fig.~\ref{fig:elasticity}b. 

Glasses were prepared by placing particles randomly on an FCC lattice, followed by running $NVE$-dynamics for 10$\tau$ where $\tau\!=\!\lambda\sqrt{m/\varepsilon}$ forms the microscopic units of time. Due to the bi-disperse nature of our model system, the FCC lattice is highly frustrated and therefore has a very large energy. The structural relaxation time of a liquid at these energies is on the order of $\tau$ or smaller. This procedure is thus equivalent to equilibrating the system at very high temperature (well above each model's glass transition temperature) liquid states. Then, each independent $NVE$-run is followed by an instantaneous quench of the system to zero temperature by means of a standard nonlinear conjugate gradient method, to form a glass. This protocol allows to compare different glass models on the same footing, as discussed at length in~\cite{boring_paper}. We created about 1000 independent glassy samples of $N\!=\!10976$ for each value of the parameter $Q$ as detailed above. Definitions of all observables considered in this work  are provided in Appendix.~\ref{app:definitions}.

\vspace{-0.2cm}

\section{Macroscopic Elastic properties}
\label{sec:results}

\begin{figure}[ht!]
  \includegraphics[width = 0.5\textwidth]{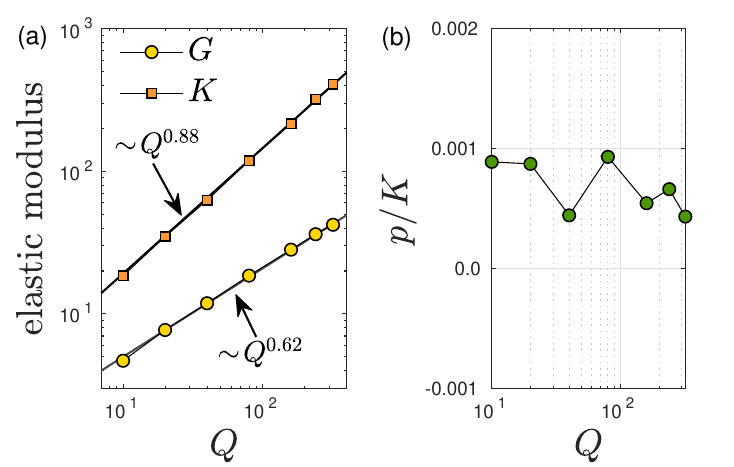}
  \caption{\footnotesize (a) The average shear ($G$) and bulk ($K$) moduli of our glass ensembles, plotted vs.~the parameter $Q$ that controls the stiffness of the repulsive part of the employed pairwise potential. (b) The densities of each $Q$-ensembles (see text for precise values) were set such that the pressure to bulk modulus ratio of all glass ensembles generated in this work are on the order of $10^{-3}$.}
  \label{fig:elasticity}
\end{figure}

We kick off the presentation with the bare macroscopic elastic properties (see definitions in Appendix~\ref{app:definitions}) of our glass ensembles. In Fig.~\ref{fig:elasticity}a we plot the raw shear and bulk moduli vs.~the parameter $Q$. These moduli follow an approximate power law as indicated in the figure. We do not attribute any particular importance to these approximate scaling laws. Interestingly, it is apparent that as $Q$ is made larger, the ratio $G/K$ tends to decrease, as also observed near the unjamming transition. Fig.~\ref{fig:elasticity}b establishes that the reduced pressure $p/K$ in our glasses is small, on the order of $10^{-3}$ for all values of the parameter $Q$ studied here, as mentioned above.

\begin{figure}[ht!]
  \includegraphics[width = 0.45\textwidth]{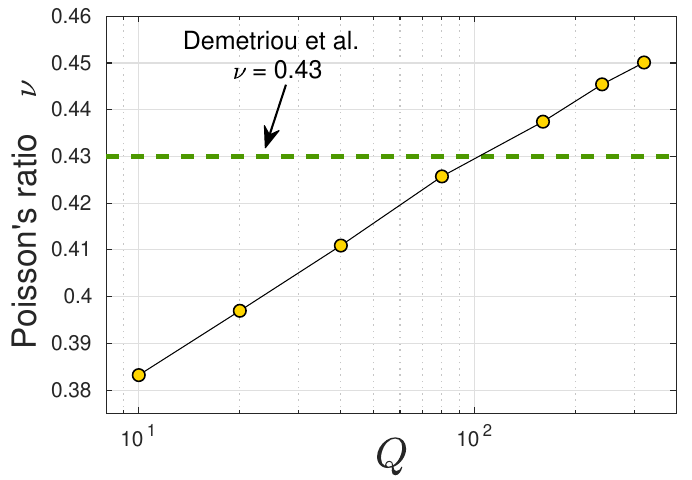}
  \caption{\footnotesize Poisson's ratio (PR) $\nu$ is plotted vs.~the parameter $Q$ that controls the interparticle stiffness (see inset) of the computer glass model put forward in this work. The dashed line corresponds to the PR of the platinum-rich glass studied in~\cite{ultra_high_PR_Demetriou_2011}, which is the highest reported for a laboratory metallic glass to date.}
  \label{fig:pr_vs_Q}
\end{figure}

In Fig.~\ref{fig:pr_vs_Q} we show the key result of this work; there the PRs $\nu\!\equiv\!\frac{3K-2G}{6K+2G}\! = \!\frac{3-2G/K}{6+2G/K}$ of our glass-ensembles are plotted against the parameter $Q$. We find that $\nu$ grows roughly logarithmically with $Q$, reaching $\nu\!=\!0.45$ for the largest $Q\!=\!320$. The dashed line represents the PR of the platinum-rich glasses reported by Demetriou et al.~in Ref.~\cite{ultra_high_PR_Demetriou_2011}, which is the largest PR reported to date (to the best of our knowledge) for a laboratory metallic glass. 

\section{Relation to the unjamming transition of repulsive soft spheres}

\subsection{Shear-to-Bulk moduli ratio}

Can the increase of $\nu$ --- or, alternatively, the \emph{decrease} in the shear-to-bulk moduli ratio $G/K$ --- in our model glass upon increasing the parameter $Q$ be understood on firmer grounds? If, for the sake of discussion, we neglect the attractive $\sim\!r^{-4}$ term in the pairwise potential (cf.~Eq.~(\ref{eq:pairwise_potential})), one would expect that $p/K\!\sim\!1/Q$, as seen in the unjamming of soft discs (in two dimensions) interacting via a purely repulsive $\propto\! r^{-Q}$ pairwise interaction potential studied in Ref.~\cite{stefanz_pre_2016}. Then, since in the unjamming scenario of repulsive spheres $G/K\!\sim\!\sqrt{p/K}$ is universally observed~\cite{stefanz_pre_2019}, we might expect that, in our glasses, $G/K\!\sim\!1/\!\sqrt{Q}$. On the other hand, the approximate power laws $G\!\sim\!Q^{0.62}$ and $K\!\sim\!Q^{0.88}$ shown in Fig.~\ref{fig:elasticity} lead us to expect that $G/K\!\sim\!Q^{-0.26}$. 

\begin{figure}[ht!]
  \includegraphics[width = 0.45\textwidth]{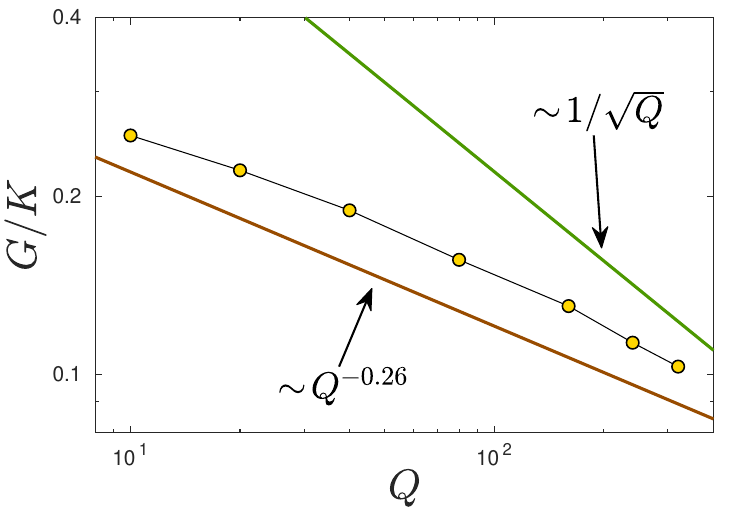}
  \caption{\footnotesize The ratio $G/K$ is plotted vs.~the parameter $Q$; the `unjamming'-based prediction $G/K\!\sim\!1/\sqrt{Q}$ (see~\cite{stefanz_pre_2016}) is not followed by our data (yellow symbols), but might be the correct asymptotic form of $G/K$ at large $Q$.}
  \label{fig:g_over_k}
\end{figure}

We test these expectations in Fig.~\ref{fig:g_over_k}. The $G/K\!\sim\!Q^{-0.26}$ expectation --- based on the approximate power laws of $G$ and $K$ in $Q$ shown above --- seems to capture the scaling of $G/K$ over a broad range of $Q$, however it appears not to be the correct asymptotic scaling of $G/K$. On the other hand, the $G/K\!\sim\!1/\!\sqrt{Q}$ scaling proposed by neglecting the role of attractive pairwise interactions is certainly not satisfied in the range of $Q$ studied here. Nevertheless, there exists a clear trend in the data indicating that $\sim\!1/\!\sqrt{Q}$ \emph{might} be the correct asymptotic form for $G/K$ at large $Q$. 

\subsection{Radial distribution functions}

\begin{figure}[ht!]
  \includegraphics[width = 0.5\textwidth]{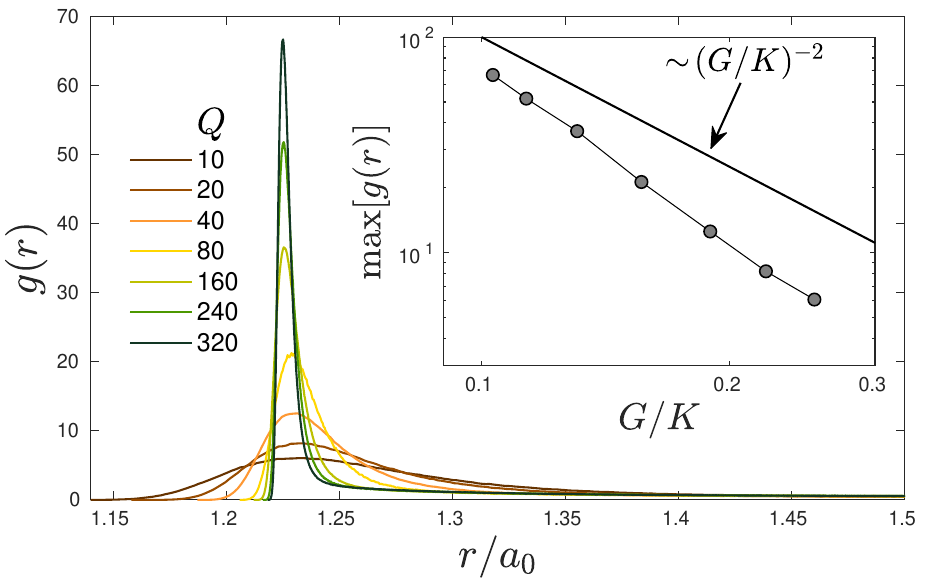}
  \caption{\footnotesize Radial distribution functions $g(r)$ of `large'-`large' particles in our model glasses for various $Q$-ensembles as specified in the legend, plotted vs.~the dimensionless distance $r/a_0$ where $a_0\!\equiv\!(V/N)^{1/3}$ is an interparticle distance. These distributions behave similarly to those measured in soft-sphere packings near unjamming~\cite{ohern2003}. Inset: the maximum of $g(r)$ increases slightly more quickly with decreasing $G/K$ than the $\sim\!(G/K)^{-2}$ scaling seen in soft-sphere unjamming.}
  \label{fig:pair_correlation}
\end{figure}

Another hallmark of the unjamming transition of soft-sphere packings is the increase in the maximal height of the first peak of the radial distribution function $g(r)$ (see definition in Appendix~\ref{app:definitions}) upon approaching the unjamming point, namely as $G/K\!\to\!0$. Previous work~\cite{ohern2003} on decompressed soft-sphere packings has established that the maximal height of the first peak grows as $(G/K)^{-2}$; in Fig.~\ref{fig:pair_correlation} we plot the maximal height of the first peak of the pair correlation function measured in our glasses with various values of the parameter $Q$. We find that the maximal height of the first peak of $g(r)$ indeed increases with increasing $Q$ (and decreasing $G/K$). However, as shown in the inset of Fig.~\ref{fig:pair_correlation}, the scaling expected from the unjamming scenario is not accurately satisfied -- we find a stronger-than-quadratic increase in the maximal height of the first peak of $g(r)$ as $G/K\!\to\!0$ over the explored $Q$-range. 

\subsection{Low-frequency vibrational spectra}

\begin{figure}[ht!]
  \includegraphics[width = 0.5\textwidth]{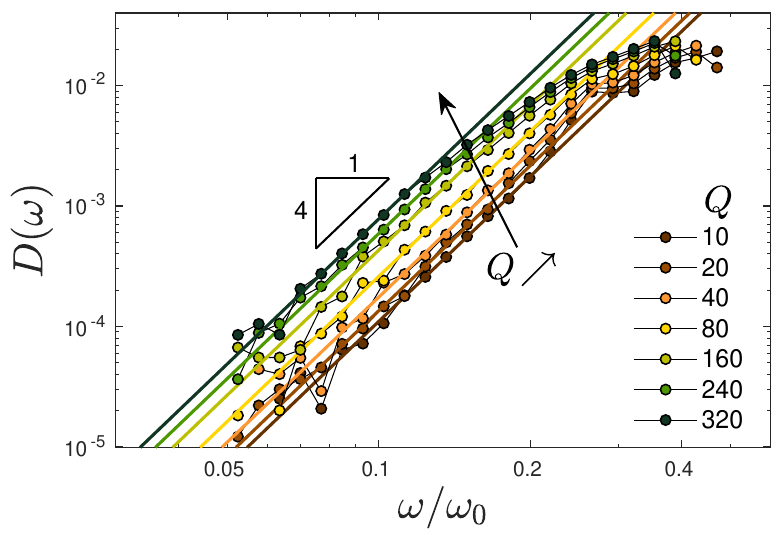}
  \caption{\footnotesize Low-frequency vibrational spectra of our glassy ensembles measured for the same $Q$ parameters as described in the main text, and plotted against the rescaled frequency $\omega/\omega_0$, where $\omega_0\!\equiv\!c_{\rm s}/a_0$ and $c_s$ is the shear-wave speed. The continuous lines represent fits to the universal $\sim\!\omega^4$ nonphononic spectra seen quite generally in structural glasses~\cite{JCP_Perspective}.}
  \label{fig:spectra}
\end{figure}

We next consider the low-frequency vibrational spectra (see precise definitions in Appendix~\ref{app:definitions}) of our glasses, displayed in Fig.~\ref{fig:spectra}. It was recently established that, below the lowest frequency shear modes in finite-size glassy samples, quasilocalized excitations populate the vibrational spectrum. Those excitations' frequencies follow a universal nonphononic distribution~\cite{JCP_Perspective} of the form $A_{\rm g}\omega^4$, with a non-universal prefactor $A_{\rm g}$ --- of dimensions [time]$^5$ --- which is indicative of the number density and characteristic frequency of quasilocalized excitations~\cite{cge_paper,pinching_pnas}. We find that the low-frequency spectra of our different $Q$-ensembles indeed follow the universal quartic law, as indicated by the low-frequency power-law fits in Fig.~\ref{fig:spectra}. Interestingly, in Refs.~\cite{ikeda_pnas,atsushi_core_size_pre} it was established that driving soft-sphere packings towards the unjamming point leads to an increasing $A_{\rm g}$~\cite{footnote3}. We also find that $A_{\rm g}$ is an increasing function of increasing $Q$ (cf.~Fig.~\ref{fig:spectra}), providing further support that increasing the parameter $Q$ in our glass-former is akin to moving closer to the unjamming transition, consistent with our working hypothesis. 

\begin{figure}[ht!]
  \includegraphics[width = 0.5\textwidth]{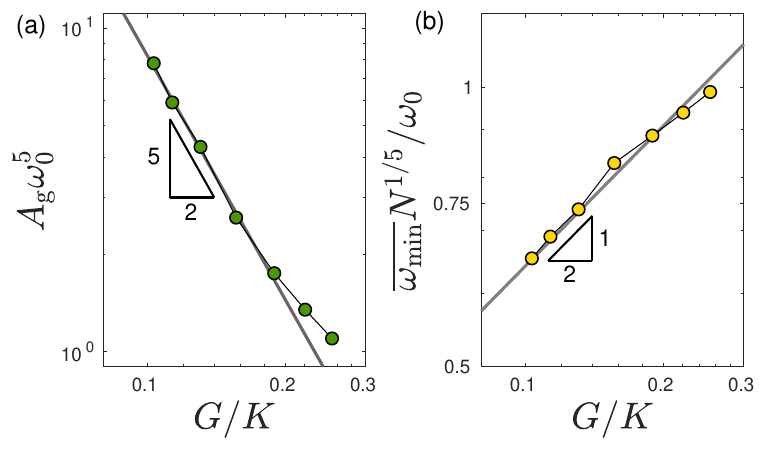}
  \caption{\footnotesize (a) The dimensionless prefactors $A_{\rm g}\omega_0^5$ of the universal nonphononic spectra of our glasses scale as $\sim\!(G/K)^{-5/2}$ for small $G/K$, as predicted from the physics of the unjamming transition. (b) The sample-to-sample mean minimal vibrational frequency of quasilocalized excitations --- made dimensionless by rescaling by $\omega_0$ --- scales as $\sim\!\sqrt{G/K}$ as predicted from the physics of the unjamming transition, see text for further discussion.}
  \label{fig:Ag_and_omega_min}
\end{figure}

What key observations can be deduced from the low-frequency spectra of our model glasses? Refs.~\cite{cge_paper,pinching_pnas} suggest and test the decomposition $A_{\rm g}\!=\!{\cal N}\omega_{\rm g}^{-5}$ where ${\cal N}$ denotes the number density of soft, quasilocalized excitations, and $\omega_{\rm g}$ is their characteristic frequency. Within the unjamming scenario of repulsive spheres the characteristic frequency $\omega_\star$ of glassy excitations is expected to follow $\omega_\star/\omega_0\!\sim\!\sqrt{G/K}$, and recall that $\omega_0\!\equiv\!c_{\rm s}/a_0$ with $c_{\rm s}$ denoting the shear wave speed and $a_0\!\equiv\!(V/N)^{1/3}$ is a typical interparticle distance. If we identify $\omega_\star$ with the characteristic frequency $\omega_{\rm g}$ of glassy quasilocalized modes~\cite{new_variational_argument_epl_2016,cge_paper}, we then expect $A_g\omega_0^5\!\sim\!(G/K)^{-5/2}$ for the dimensionless prefactors of the universal nonphononic spectra. Good asymptotic agreement with this expectation is shown in Fig.~\ref{fig:Ag_and_omega_min}a.

Another way to probe the scaling properties of glassy excitations' characteristic frequency $\omega_{\rm g}$ is via the sample-to-sample mean of the minimal vibrational frequency $\overline{\omega_{\rm min}}$. In~\cite{modes_prl_2016} it was shown that the sample-to-sample statistics of the minimal frequencies of quasilocalized modes is Weibullian. In particular, their sample-to-sample means follow $\overline{\omega_{\rm min}}\!\sim\!\omega_g N^{-1/5}$ by virtue of the universal $\sim\!\omega^4$ nonphononic spectrum. We thus expect $\overline{\omega_{\rm min}}/\omega_0\!\sim\!\sqrt{G/K}$, as indeed validated in Fig.~\ref{fig:Ag_and_omega_min}b.


\section{Mesoscopic mechanical disorder}

Having considered the behavior of the microscopic structure ($g(r)$) and of the properties of micromechanical excitations, we now turn to assessing the degree of mesoscopic mechanical disorder of our glass ensembles, and relating it to the expectations from the unjamming scenario. To this aim, we consider the sample-to-sample standard-deviation to mean ratio of the shear modulus $G$, factored by $\sqrt{N}$ to obtain an $N$-independent mesoscopic mechanical disorder quantifier, denoted as $\chi$ in what follows, namely
\begin{equation}
    \chi \equiv \frac{\sqrt{N\overline{(G - \overline{G})^2}}}{\overline{G}}\,,
\end{equation}
where $\overline{\bullet}$ stands for an ensemble average. The quantifier $\chi$ has been shown to control wave attenuation rates in the harmonic and long-wavelength regime~\cite{scattering_letter_jcp_2021}, and also finds manifestations in finite-size effects of the vibrational spectra of disordered solids~\cite{phonon_widths,phonon_widths2}.

\begin{figure}[ht!]
  \includegraphics[width = 0.5\textwidth]{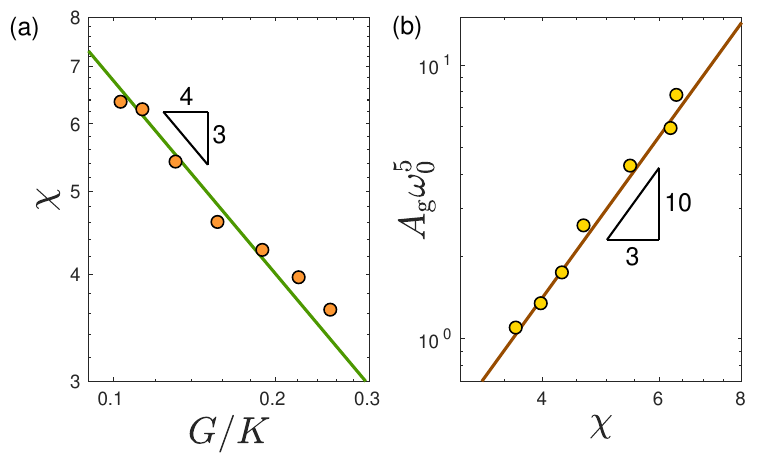}
  \caption{\footnotesize (a) The mechanical disorder quantifier $\chi$ is plotted against the ratio $G/K$. The agreement with prediction $\chi\!\sim\!(G/K)^{-3/4}$ spelled out in the text is reasonable; larger statistical ensembles and larger glassy samples might reduce the noise in $\chi$ and result in better agreement. (b) The dimensionless prefactors $A_{\rm g}\omega_0^5$ --- extracted from the spectral analysis presented in Fig.~\ref{fig:spectra} --- are plotted against the quantifier $\chi$ of mesoscopic mechanical disorder. The continuous line represents the prediction $A_{\rm g}\omega_0^5\!\sim\!\chi^{10/3}$ spelled out and tested in Ref.~\cite{david_fracture_2021}, and also explained in the text.}
  \label{fig:chi_vs_Ag}
\end{figure}

In Fig.~\ref{fig:chi_vs_Ag}a we plot the disorder quantifier $\chi$ vs.~the ratio $G/K$. To rationalize our observations, we assume that $\chi^2$ represents a dimensionless correlation \emph{volume}~\cite{bouchaud_random_walks_1990} of the shear modulus spatial fluctuations, and we build upon the observation~\cite{pinching_pnas} that the linear size $\xi_{\rm g}$ of quasilocalized excitations (see discussion in~\cite{JCP_Perspective}) is inversely proportional to their characteristic frequency, namely $\xi_{\rm g}/a_0\!\sim\!\omega_0/\omega_{\rm g}$. If the size of soft quasilocalized excitations represents the mesoscopic correlation volume of the shear modulus spatial fluctuations, one expects $\chi^2\!\sim\!(\xi_{\rm g}/a_0)^3\!\sim\!(\omega_0/\omega_{\rm g})^{-3}$ (in three dimensions). Finally, since $\omega_{\rm g}/\omega_0\!\sim\!\sqrt{G/K}$ (cf.~Fig.~\ref{fig:Ag_and_omega_min}b and accompanying discussion), we expect $\chi\!\sim\!(G/K)^{-3/4}$. This prediction is superimposed onto the data in Fig.~\ref{fig:chi_vs_Ag}a, to find reasonable agreement. 

Can properties of the micromechanical excitations in our model glasses be related to the latters' mesoscopic elastic properties? To establish such a relation, we further plot in Fig.~\ref{fig:chi_vs_Ag}b the dimensionless prefactors $A_{\rm g}\omega_0^5$ vs.~the mechanical disorder quantifier $\chi$. Following the same line of reasoning spelled out in the preceding paragraphs, together with the scaling $A_{\rm g}\!\sim\!\omega_g^{-5}$ (also established in~\cite{sticky_spheres2}) and similarly to the results presented in~\cite{sticky_spheres1_karina_pre2021,david_fracture_2021}, we find good agreement with the prediction $A_{\rm g}\omega_0^5\!\sim\!\chi^{10/3}$ relating between two independently-measured microscopic ($A_{\rm g}\omega_0^5$) and mesoscopic ($\chi$) quantifiers of mechanical disorder.

We end this Section by noting that, within the unjamming scenario of relaxed disordered spring networks of mean coordination $z$, one expects $\chi\!\sim\!1/\sqrt{z\!-\!z_c}$ --- as shown in~\cite{phonon_widths2} ---, where $z_c\!\equiv\!2\dbar$ is the Maxwell threshold, and $\dbar$ denotes the dimension of space. Based on this result, one expects that in the unjamming of repulsive spherical particles $G/K\!\sim\!z\!-\!z_c$ and thus $\chi\!\sim\!1/\sqrt{G/K}$, at odds with our data presented in Fig.~\ref{fig:chi_vs_Ag}a. However, preliminary data (not shown) indicates that the scaling $\chi\!\sim\!1/\sqrt{z\!-\!z_c}$ in relaxed disordered spring networks only holds when the correlation length $\xi_{\rm g}/a_0\!\gg\!1$, a condition not yet satisfied by our glasses. We therefore speculate that 
\begin{equation}
\chi  \sim \left\{ 
\begin{array}{cc}
(\xi_{\rm g}/a_0)^{\dbar/2} \,,    & \xi_{\rm g}/a_0\!\sim\!{\cal O}(1)  \\
(\xi_{\rm g}/a_0)^{1/2}\,,     &  \xi_{\rm g}/a_0\!\gg\!1 
\end{array}\right.\,,
\end{equation}
with a crossover taking place at around $\xi_{\rm g}/a_0\!\approx\!10$. Future research should validate or refute this speculation.

\section{Softening mechanism of low-energy excitations}

We end the discussion by highlighting an interesting \emph{qualitative} difference between the model glass introduced and studied here, and soft-sphere packings near the unjamming transition. In particular, in soft-sphere packings, the main softening mechanism of low-energy excitations originates from the destabilizing effect of compressive interparticle forces that lead to buckling-like instabilities and associated emergent soft excitations~\cite{matthieu_PRE_2005,eric_boson_peak_emt,inst_note}. In contrast, in our ultra-high PR model glasses, soft-mode energies are reduced predominantly due to the utilization of the interparticle-potential's negative \emph{stiffnesses} (see related discussion in~\cite{sticky_spheres1_karina_pre2021}) -- a mechanism that is entirely absent in decompressed, purely repulsive packings near the unjamming point.

To be more quantitative, following Ref.~\cite{sticky_spheres1_karina_pre2021} we define the following decomposition of the Hessian matrix:
\begin{eqnarray}
    \calBold{H} & \equiv & \frac{\partial^2 U}{\partial\xv\partial\xv} = \sum_{i<j}\varphi_{ij}''\frac{\partial r_{ij}}{\partial\xv}\otimes\frac{\partial r_{ij}}{\partial\xv} + \sum_{i<j}\varphi'_{ij}\frac{\partial^2r_{ij}}{\partial\xv\partial\xv} \nonumber \\
    & = & \sum\limits_{\varphi_{ij}''>0}\varphi_{ij}''\frac{\partial r_{ij}}{\partial\xv}\otimes\frac{\partial r_{ij}}{\partial\xv} + \sum\limits_{\varphi_{ij}''<0}\varphi_{ij}''\frac{\partial r_{ij}}{\partial\xv}\otimes\frac{\partial r_{ij}}{\partial\xv} \nonumber \\
    & &\ \  + \sum\limits_{\varphi_{ij}'>0}\varphi'_{ij}\frac{\partial^2r_{ij}}{\partial\xv\partial\xv} + \sum\limits_{\varphi_{ij}'<0}\varphi'_{ij}\frac{\partial^2r_{ij}}{\partial\xv\partial\xv} \nonumber \\
    & \equiv & \ \  \calBold{H}''_+ +  \calBold{H}''_- + \calBold{H}'_+ + \calBold{H}'_- \,, \label{eq:argument}
\end{eqnarray}
where $\calBold{H}''_+,\calBold{H}'_+$ are positive definite, $\calBold{H}''_-,\calBold{H}'_-$ are negative definite, and $\otimes$ denotes an outer product. With the above decomposition of $\calBold{H}$, the energy $\omega^2\!=\!\piv\cdot\!\calBold{H}\!\cdot\piv>\!0$ associated with any given, translation-free mode $\piv$ can be written as
\begin{equation}\label{eq:frequency_decomposition}
    \omega^2 = k_+ + k_- + f_+ + f_- \,,
\end{equation}
where 
\begin{eqnarray}
k_+ & \equiv & \piv\cdot\calBold{H}''_+\cdot\piv\,,\quad\quad k_-  \equiv  \piv\cdot\calBold{H}''_-\cdot\piv\,, \nonumber \\
f_+ & \equiv & \piv\cdot\calBold{H}'_+\cdot\piv\,,\quad\quad f_-  \equiv  \piv\cdot\calBold{H}'_-\cdot\piv\,, \nonumber  
\end{eqnarray}
and we note that $k_+,k_-,f_+,f_-$ all have units of energy/length$^2$, assuming $\piv$ is normalized and thus dimensionless. We further note that, in purely repulsive soft-sphere packings, $f_+\!=\!k_-\!=\!0$, however these terms are generally nonzero once attractive interactions are introduced between the constituent particles. 


In Fig.~\ref{fig:chi_vs_Ag}b we report the ratio $k_-/f_-$ for soft quasilocalized excitations analyzed using the nonlinear framework discussed in~\cite{episode_1_2020}, for our various $Q$-ensembles (see Appendix~\ref{app:definitions} for definitions and details). As mentioned above, we find that, for the highest PR glasses, the destabilizing terms pertaining to negative stiffness, $k_-$, dominate over the buckling-like terms ($f_-$) in softening the lowest-energy excitations. Interestingly, the precise mechanism that softens low-energy excitations has no effect on the functional form of the VDoS of structural glasses~\cite{sticky_spheres1_karina_pre2021,modes_prl_2020}, as also demonstrated here. The signatures of this different softening mechanism on the anatomy and geometry of plastic instabilities under external loading~\cite{avraham_core_properties_pre_2020} is left for future investigations.

\begin{figure}[ht!]
  \includegraphics[width = 0.45\textwidth]{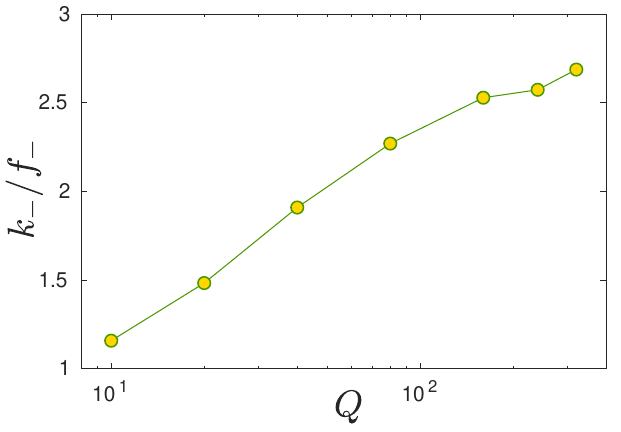}
  \caption{\footnotesize The ratio between the destabilizing stiffness and force terms, $k_-/f_-$, measured for low-energy quasilocalized excitations, and plotted against the parameter $Q$. These data highlight a qualitative difference in the micromechanics of our glasses compared to repulsive sphere-packings for which $k_-/f_-$ vanishes identically.}
  \label{fig:stiffness_ratio}
\end{figure}

\section{Summary and outlook}
\label{sec:summary}

Inspired by experimental work on platinum-rich glasses~\cite{ultra_high_PR_Demetriou_2011}, in this work we raise and test the hypothesis that the mechanism controlling the Poisson's ratio in highly incompressible glasses is related to the unjamming transition of packings of soft repulsive spheres. To this aim we design a glass forming model --- with a finite tensile yield-stress --- whose repulsive stiffness can be tuned via a single parameter $Q$, and investigate the model's elastic and vibrational properties under variations of $Q$. We find that at large $Q$ our glasses feature ultra-high PRs, exceeding the most incompressible laboratory metallic glasses.

In addition, we highlight several unjamming-like phenomena observed in our model glasses; in particular, we find an increase in the first peak of the pair-correlation function $g(r)$, an increase of the prefactor $A_{\rm g}$ of the $\sim\!\omega^4$ universal frequency distribution of nonphononic, quasilocalized excitations~\cite{ikeda_pnas,atsushi_core_size_pre}, and a decrease in the relative characteristic frequency of those excitations~\cite{atsushi_core_size_pre}, upon increasing the interparticle interaction stiffness (via the parameter $Q$). Finally, we point out an interesting, qualitative difference between our ultra-high PR glasses and soft-sphere packings in terms of the micromechanics of their respective soft, quasilocalized excitations: in our glasses the main softening mechanism is not micromechanical buckling --- as it is in repulsive sphere packings ---, but rather via exploiting the negative (unstable) stiffnesses associated with the \emph{attractive} part of the pairwise potential between the constituent particles. 

Interestingly, not only does the platinum-rich glass studied in~\cite{ultra_high_PR_Demetriou_2011} feature an ultra-high PR, but it also features a very high bulk modulus (of $\approx$~217 GPa), consistent with the mechanism discussed here that involves increasing the stiffness of the pairwise interaction potential of our model systems to obtain high-PR glasses. Future research should focus on more stringent tests of the unjamming-like picture described in this work, and its relevance to ultra-high PR laboratory glasses.

We further highlight that increasing the stiffness of pairwise potentials in models of disordered solids does not generally lead to higher PRs in the resulting glasses. For example, in~\cite{riggleman_ductile_to_brittle_soft_matter_2019} it was shown that increasing the stiffness of the pairwise potential of a model glass can lead to a sharp ductile-to-brittle transition in the corresponding nonlinear mechanical response of the resulting glasses. Since ductile-to-brittle transitions are known to be correlated to a \emph{decreasing} PR~\cite{Greer_2005,Greaves2011}, one concludes that stiffening the pairwise potential can also lead to the \emph{decrease} in the PR. We speculate that the essential difference between the observations of~\cite{riggleman_ductile_to_brittle_soft_matter_2019} and the present work the have to do with the strength of nonaffinity under compressive strains (see lengthy discussion in~\cite{stefanz_pre_2019}), which is suppressed in the model presented here. Since enhanced nonaffinity leads to the softening of the bulk modulus, it consequently results in a \emph{lower} PR of the corresponding glasses. Establishing these aforementioned speculations is left for future work.

Finally, we highlight that the independent validation of the scaling laws $A_g\omega_0^5\!\sim\!(G/K)^{-5/2}$ and $\omega_g/\omega_0\!\sim\!\sqrt{G/K}$ puts on firmer grounds our assumption that the number density ${\cal N}$ of quasilocalized excitations only weakly depends on $Q$ in our glass models. It remains to be seen to what extend thermal annealing can deplete the number density of quasilocalized excitations near the unjamming transition.



\acknowledgements
We warmly thank Eran Bouchbinder for insightful discussions. Support from the NWO (Vidi grant no.~680-47-554/3259) is acknowledged.

\appendix


\section{Definitions of observables}
\label{app:definitions}

The observables we consider in this work are detailed in this Appendix.

We start with athermal elastic moduli \cite{lutsko}; the shear modulus $G$ is defined as
\begin{equation}\label{eq-G}
    G \equiv \frac{1}{V}\frac{d^2U}{d\gamma^2}= \frac{\frac{\partial^{2}U}{\partial \gamma^{2}}-\frac{\partial^{2}U}{\partial \gamma\partial \xv} \cdot \calBold{H}^{-1}\cdot \frac{\partial ^{2}U}{\partial\xv\partial \gamma}}
    {V}\,,
\end{equation}
where $d/d\gamma$ denotes the total derivative under the constraints of mechanical equilibrium at zero temperature \cite{lutsko}, $\xv$ denotes particles' coordinates, $\calBold{ H}\!\equiv\!\frac{\partial^2U}{\partial\xv\partial\xv}$ is the Hessian matrix of the potential energy $U\!=\!\sum_{i<j}\varphi_{ij}$, $\varphi_{ij}$ is the pairwise interaction potential (see Eq.~(\ref{eq:pairwise_potential})) and $\gamma$ is a shear-strain parameter that parameterizes the imposed affine simple shear (in the $x$-$y$ plane) transformation of coordinates $\xv\!\to \mathBold{H}(\gamma)\cdot\xv$ with
\begin{equation}\label{shear_transformation_matrix}
\mathBold{H}(\gamma) =  \left( \begin{array}{ccc}1&\gamma&0\\0&1&0\\
0&0&1\end{array}\right)\,.
\end{equation}

The bulk modulus $K$ is defined as
\begin{equation}\label{eq-K}
    K \equiv -\frac{1}{\dbar}\frac{dp}{d\eta} = \frac{\frac{\partial^{2}U}{\partial \eta^{2}}\! -\!\dbar \frac{\partial U}{\partial\eta}\! -\! \frac{\partial^{2}U}{\partial \eta \partial \xv}\! \cdot\! \calBold{H}^{-1}\! \cdot\! \frac{\partial^{2}U}{\partial\xv\partial\eta}}{V\dbar^2}\,,
\end{equation}
where 
\begin{equation}
p\equiv-\frac{1}{V\dbar}\frac{\partial U}{\partial \eta}
\end{equation}
is the pressure, $\dbar$ is the dimension of space, and $\eta$ is an expansive-strain parameter that parameterizes the imposed affine expansive transformation of coordinates $\xv\!\to \mathBold{H}(\eta)\cdot\xv$ as
\begin{equation}\label{dilation_transformation_matrix}
\mathBold{H}(\eta) =  \left( \begin{array}{ccc}e^\eta&0&0\\0&e^\eta&0\\0&0&e^\eta\end{array}\right)\,.
\end{equation}
With the definitions of the shear and bulk moduli in hand, the Poisson's ratio $\nu$ of a three-dimensional solid is given by
\begin{equation}
    \nu \equiv \frac{3K-2G}{6K+2G} = \frac{3-2G/K}{6+2G/K}\,.
\end{equation}
The speed of shear waves is given by $c_{\rm s}\!\equiv\!\sqrt{G/\rho}$ where $\rho$ denotes the mass density.

We consider vibrational modes' frequencies $\omega$ obtained via the eigenvalue equation
\begin{equation}
\label{eigenvalue_equation}
    \calBold{H}\cdot\psiv_{\omega} = \omega^2\psiv_\omega\,,
\end{equation}
where $\psiv_\omega$ denotes the eigenvector pertaining to the eigenvalue $\omega^2$, and we note that all masses are set to unity. The eigenfrequencies are histogram to obtain the glasses' spectra $D(\omega)$ shown in Fig.~\ref{fig:spectra}. 

The properties of low-energy quasilocalized excitations are studied using the nonlinear framework discussed e.g.~in~\cite{episode_1_2020}. Within this framework, quasilocalized excitations are given by solutions $\piv$ to the equation
\begin{equation}\label{eq:cubic_modes}
    \calBold{H}\cdot\piv = \frac{\calBold{H}:\piv\piv}{{\bm U'''}\tripleCdot\piv\piv\piv}{\bm U'''}:\piv\piv\,,
\end{equation}
where ${\bm U'''}\!\equiv\!\frac{\partial U}{\partial\xv\partial\xv\partial\xv}$ denotes the rank-3 tensor of derivatives of the potential energy, and $\cdot,:,\tripleCdot$ denote single, double or triple contractions over particle indices and Cartesian components, respectively. In~\cite{episode_1_2020}, it was shown that excitations $\piv$ defined via Eq.~(\ref{eq:cubic_modes}) closely resemble quasilocalized vibrational (harmonic) modes, in the absence of hybridizations with low-frequency phonons. Solutions $\piv$ were calculated as explained in Appendix~B of~\cite{episode_1_2020}, which presumably provides one of the lowest-energy glassy excitations in a given glass sample. 

Finally, we also considered the radial distribution function $g(r)$ defined as
\begin{equation}
    g(r) \equiv \frac{2V}{N^2}\bigg< \sum_{i<j}\delta(r-r_{ij})\bigg>\,,
\end{equation}
where $r_{ij}\!\equiv\!\sqrt{(\xv_j\!-\!\xv_i)\cdot(\xv_j\!-\!\xv_i)}$ is the pairwise distance between particles $i$ and $j$.



\begin{thebibliography}{50}
\bibitem [{foo()}]{footnote3}%
  \BibitemOpen
  \bibinfo {note} {In Ref.~\cite{atsushi_core_size_pre} it is argued that
  $A_{\rm g}\omega_0^5\!\sim\!(G/K)^{-3/2}$, however we find a stronger
  scaling, of $A_{\rm g}\omega_0^5\!\sim\!(G/K)^{-5/2}$, as shown in
  Fig.~\ref{fig:Ag_and_omega_min}a.}\BibitemShut {Stop}%
\end{thebibliography}
%

\end{document}